\documentclass{article}

\usepackage[preprint]{neurips_2024}

\usepackage[utf8]{inputenc} % allow utf-8 input
\usepackage[T1]{fontenc}    % use 8-bit T1 fonts
\usepackage{hyperref}       % hyperlinks
\usepackage{url}            % simple URL typesetting
\usepackage{booktabs}       % professional-quality tables
\usepackage{amsfonts}       % blackboard math symbols
\usepackage{nicefrac}       % compact symbols for 1/2, etc.
\usepackage{microtype}      % microtypography
\usepackage[svgnames]{xcolor}         % colors
\usepackage{wrapfig}
\usepackage{listings}
\definecolor{codegreen}{rgb}{0,0.6,0}
\definecolor{codegray}{rgb}{0.5,0.5,0.5}
\definecolor{codepurple}{rgb}{0.58,0,0.82}
\definecolor{backcolour}{rgb}{0.95,0.95,0.92}

\lstdefinestyle{mystyle}{
    backgroundcolor=\color{backcolour},   
    commentstyle=\color{codegreen},
    keywordstyle=\color{magenta},
    numberstyle=\tiny\color{codegray},
    stringstyle=\color{codepurple},
    basicstyle=\ttfamily\footnotesize,
    breakatwhitespace=false,         
    breaklines=true,                 
    captionpos=b,                    
    keepspaces=true,                 
    numbers=left,                    
    numbersep=5pt,                  
    showspaces=false,                
    showstringspaces=false,
    showtabs=false,                  
    tabsize=2
}
\usepackage{circledsteps}\pgfkeys{/csteps/fill color=FireBrick, /csteps/outer color=FireBrick, /csteps/inner color=white}

\usepackage{subcaption}

\usepackage{graphicx}       % figures
\RequirePackage{here}
\graphicspath{{figures/}{./}} 

\title{The Unreasonable Effectiveness of LLMs \\ for Query Optimization}

\author{%
  Peter Akioyamen%\thanks{Use footnote for providing further information
    \\
  University of Pennsylvania\\
  \texttt{peterai@seas.upenn.edu} 
  \And 
    Zixuan Yi %\thanks{Use footnote for providing further information
    \\
  University of Pennsylvania\\
  \texttt{zixy@seas.upenn.edu}
  \And 
    Ryan Marcus %\thanks{Use footnote for providing further information
    \\
  University of Pennsylvania\\
  \texttt{rcmarcus@seas.upenn.edu}
}

\begin{document}

\maketitle

\begin{abstract}
Recent work in database query optimization has used complex machine learning strategies, such as customized reinforcement learning schemes. Surprisingly, we show that LLM embeddings of query text contain useful semantic information for query optimization. Specifically, we show that a simple binary classifier deciding between alternative query plans, trained only on a small number of labeled embedded query vectors, can outperform existing heuristic systems. Although we only present some preliminary results, an LLM-powered query optimizer could provide significant benefits, both in terms of performance and simplicity.
\end{abstract}

\section{Introduction}

Query optimization is the task of transforming complex SQL queries into efficient programs~(\cite{systemr}), referred to as query plans. Optimizers represent substantial engineering efforts~(\cite{testing_mssql}), often spanning hundreds of thousands of lines of code~(\cite{volcano}). Most query optimizers today are driven by complex, manually-written heuristics. Despite significant advancements, query optimizers (QOs) are far from perfect, frequently making costly mistakes~(\cite{leis2015good}). 

Recent work has shown that machine learning techniques can be used to \textit{steer} query optimizers in the right direction, helping the optimizer determine which plan to select for query execution. Researchers have used supervised learning (e.g.,~\cite{fastgres}), reinforcement learning (e.g., ~\cite{marcusbao}), and hybrid approaches (e.g.,~\cite{anneser2023autosteer}) to effectively steer optimizers. However, each approach performs sophisticated feature engineering on statistics kept internally by the database, and, as a result, requires complex and deep integration with the underlying query optimizer. This has a number of downsides that have hindered practical adoption~(\cite{pilotscope}). 

In this extended abstract, we present initial results for \textsc{LLMSteer}, a  simpler approach to steering QOs. Instead of manually engineering complex features from plans or data statistics, we use a large language model (LLM) to embed raw SQL submitted by the database user. We then train a supervised learning model on a small labeled set of queries to predict the optimal direction in which to steer the QO. This places the entire ``steering'' component outside of the database, simplifying integration. 

\begin{figure}[h]
    \centering
    \includegraphics[width=\linewidth]{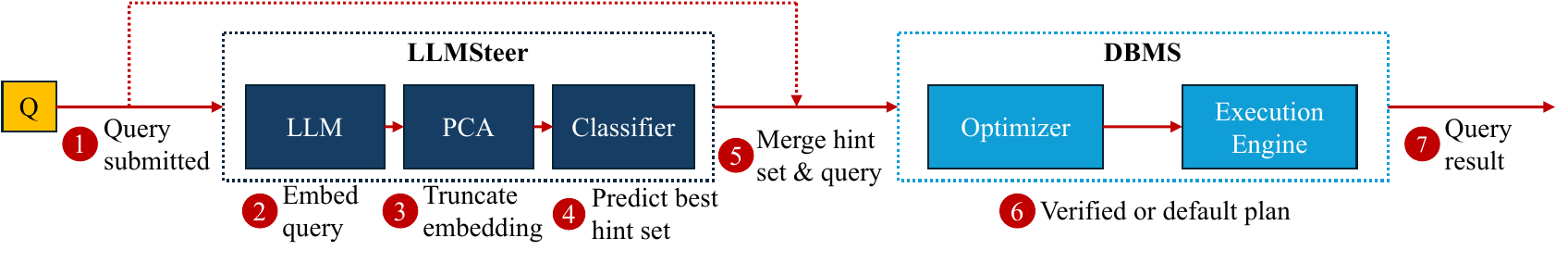}
    
    \caption{\textsc{LLMSteer} system model.}
    \label{fig:llmsteer}
\end{figure}

\textbf{As database experts, we did not expect this simple approach to work.} Common wisdom within the database community is that complex features --- such as cardinality estimates~(\cite{kipf2018learned}) or operator models~(\cite{Heinrich_2022}) --- are required for the task. Experimentally, we show that LLMs are capable of making these decisions without any such information. We found no simple explanation for LLMs' apparent success; the LLM-based approach was insensitive to at least some syntax changes, and worked across two different workloads. In Section~\ref{sec:LLMSteer}, we describe \textsc{LLMSteer} and its simple, yet powerful, design. In Section~\ref{sec:results}, we present results from initial experiments. We conclude in Section~\ref{sec:conclusion} with a discussion of future directions and questions left unanswered. 

\section{\textsc{LLMSteer}: A Surprisingly Simple Approach to Query Steering} \label{sec:LLMSteer}

\paragraph{Query hints} Given a SQL query, an optimizer can generate several plan variants, each of which may use different operators or data access patterns. \emph{Hints} are optional keywords or clauses that can be inserted into a query to guide the optimizer into generating plans with specific characteristics, providing a coarse-grained way to influence a query's execution plan. For example, a hint may indicate to the optimizer that it should only consider plans with hash joins, make use of a helpful index, or limit parallelism. 

\paragraph{Optimizer steering} Of course, determining the correct hint for a query requires \emph{a priori} knowledge of the data and workload. \emph{Steering} an optimizer is the task of selecting a hint or set of hints ("hint set") for a particular query such that the selected plan results in reduced or minimal latency. Although hints can be effective in fine-tuning database performance, selecting hints can be extremely complicated, and providing the optimizer with incorrect hints can severely degrade query latency. As a result, the practice of manually issuing a hint is used sparingly, and is commonly restricted to experts most familiar with the underlying data~(\cite{marcusbao}).

\paragraph{Problem definition} \textsc{LLMSteer} attempts to automatically determine an appropriate hint for a query once that query is submitted to the system. Given a small set of labeled embedded query vectors, \textsc{LLMSteer} trains a classification model to map unseen queries to an appropriate hint. We evaluate the quality of \textsc{LLMSteer}'s decisions using two common metrics (\cite{redshift_workload}): the change in total and P90 tail latency. When executing a query workload, \emph{total latency} is the cumulative execution time of all queries. \emph{P90 tail latency} is the 90th percentile query latency.  

\paragraph{\textsc{LLMSteer}} An overview of \textsc{LLMSteer} is depicted in Figure~\ref{fig:llmsteer}. When a query is first submitted to \textsc{LLMSteer} (\Circled{1}), the raw SQL is embedded using a large language model, producing an embedding vector (\Circled{2}). Since our goal is to train a supervised learning model using a small number of examples $k$, and since LLMs often use high $d$-dimensional embeddings (i.e., $d > k$), we next apply dimensionality reduction (\Circled{3}). The final feature vector is passed through a classifier to determine the choice of optimal hint for the given query (\Circled{4}). The hint is then combined with the original SQL query (\Circled{5}) and submitted to the database management system (DBMS) where a query plan is generated and executed (\Circled{6}).

\begin{figure}[h] 
    \centering
    \includegraphics[width=0.75\linewidth]{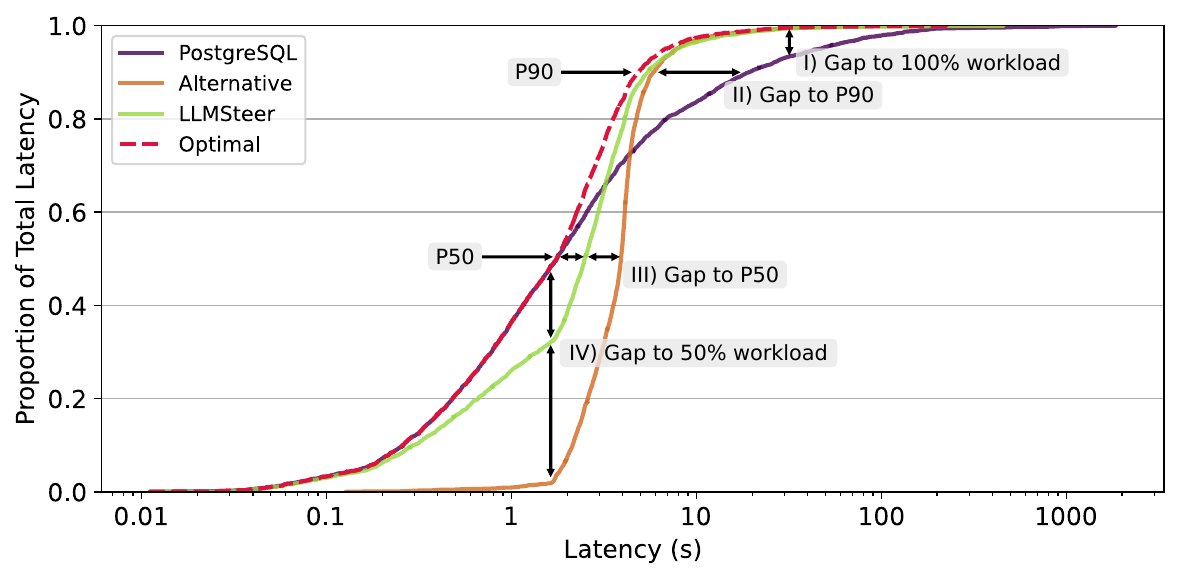}
    
    \caption{Empirical CDF of latency across cross-validation testing workloads by steering strategy. Note that the latency axis is set to $\log$ scale.}
    \label{fig:ecdf}
\end{figure}

\section{Initial Results} \label{sec:results}

We set out to answer two preliminary questions: first, can \textsc{LLMSteer} find hints for queries that outperform existing query optimizers? Second, since \textsc{LLMSteer} operates on SQL syntax directly, how robust or sensitive is \textsc{LLMSteer} to non-semantic syntactic changes in SQL queries (e.g., changes in indentation or whitespace)?

\paragraph{Experimental setup} Data used in this work contains 3246 SQL queries, 113 originating from \cite{leis2015good}'s Join Order Benchmark (JOB)  and the remaining 3133 are the "core" subset of \cite{negi2021flow}'s Cardinality Estimation Benchmark (CEB).\footnote{The "core" subset was specially selected to minimize similarity and overlap in query structure.} To obtain reliable estimates of latencies, each query is executed under a given hint set 5 times and latencies are averaged over the runs to yield a final latency that is used (\cite{10.1145/3663742.3663974}). All queries are executed using PostgreSQL version 16.1. A collection of 48 hint sets is considered, the same as those used by \cite{marcusbao} and \cite{Heinrich_2022}. To generate embeddings, we use OpenAI's \texttt{text-embedding-3-large} model. Given the latencies of queries under each hint, we determine the hint with the highest potential for improvement if applied perfectly, and we use this hint as our alternative plan (i.e., \textsc{LLMSteer} decides between the default plan and the alternative plan produced by giving the selected hint to the DBMS). Afterward, a binary label is generated for each query by determining which of the two possibilities (the default plan or the alternative plan) produces the query plan with lower latency. Approximately 30\% of queries in the data perform better with the alternative hint. 

We considered a number of models and found support vector machines with the RBF kernel trained on 120 principal components most performant; models trained on 5, 50, and 120 principal components were evaluated, preserving approximately 50\%, 80\%, and 90\% of variance in the original embeddings. Class weights were used, defined as the ratio of class frequencies, otherwise no hyperparameter tuning was performed and default values were used for all models --- the final SVM model used regularization strength $C=1.0$ and a kernel coefficient of $\gamma=1/120\sigma^2_X$. To train models, we employed a 10-fold cross-validation procedure with stratified random sampling. Our code, along with the data used in the analysis and produced embeddings, are available on GitHub.\footnote{\href{https://github.com/peter-ai/LLMSteer}{https://github.com/peter-ai/LLMSteer}}

\begin{figure}[h] 
    \centering
        \begin{subfigure}[b]{0.49\linewidth} 
            \centering
            \includegraphics[width=\linewidth]{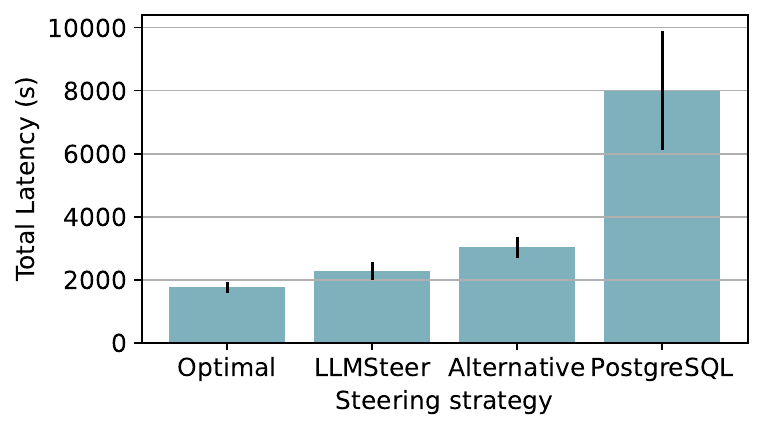}
            
            \caption{Total latency.}
            \label{fig:sum}
        \end{subfigure}
        \hspace{3.25pt}   % Space between the figures
        \begin{subfigure}[b]{0.49\linewidth}
            \centering
            \includegraphics[width=\linewidth]{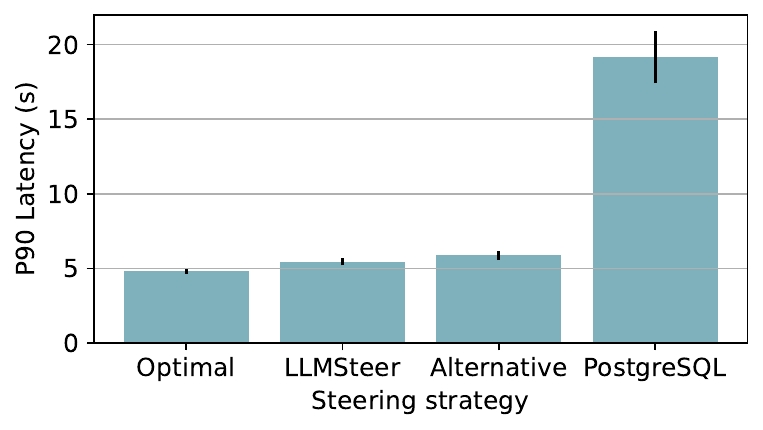}
            
            \caption{P90 latency.}
            \label{fig:p90}
        \end{subfigure}
        
    \caption{Mean \textsc{LLMSteer} performance on 10-fold cross-validation testing workloads.}
    \label{fig:performance}
\end{figure}

\paragraph{Experiment 1: \textsc{LLMSteer} vs. PostgreSQL Optimizer} Figure~\ref{fig:ecdf} shows the distribution of query latencies using four different strategies: the optimal (unknowable in practice), the default plan, the alternative plan, and \textsc{LLMSteer}. \textsc{LLMSteer} falls short of the optimal steering strategy, but effectively combines the benefits of PostgreSQL and the alternative hint. PostgreSQL's default plan outperforms the alternative plan at the lower end of the latency distribution, highlighted by PostgreSQL having a lower median latency and reaching 50\% of total workload latency earlier in the empirical CDF (IV). These dynamics invert at the higher end of the distribution, where the alternative has a significantly lower P90 (II). \textsc{LLMSteer} tracks the performance of the optimal strategy in the latter half of the latency distribution, saturating faster than both PostgreSQL and the alternative (I). The system also has a smaller performance gap to PostgreSQL and the optimal relative to the alternative plan earlier in the latency distribution, achieving a lower median latency on testing workloads (III). Thus, \textsc{LLMSteer} can be seen as trading a small increase in median latency for a large reduction in P90 and total latency, a tradeoff many practitioners would eagerly accept~(\cite{redshift_workload}).

We evaluate the performance of \textsc{LLMSteer} against the native PostgreSQL optimizer on P90 and total latency in Figure~\ref{fig:performance} as well. \textsc{LLMSteer} represents a significant improvement on the PostgreSQL default, reducing total and P90 latency by 72\% on average across testing cross-validation folds. \textsc{LLMSteer} performs near optimal relative to the steering strategy that selects the correct hint set for every query, achieving a total and P90 latency that is only 30\% and 12\% higher. Additionally, \textsc{LLMSteer} shows stability, with performance gains consistent across testing workloads with minimal deviation. 

\begin{wrapfigure}{r}{0.44\textwidth}
    \centering
    \includegraphics[width=0.44\textwidth]{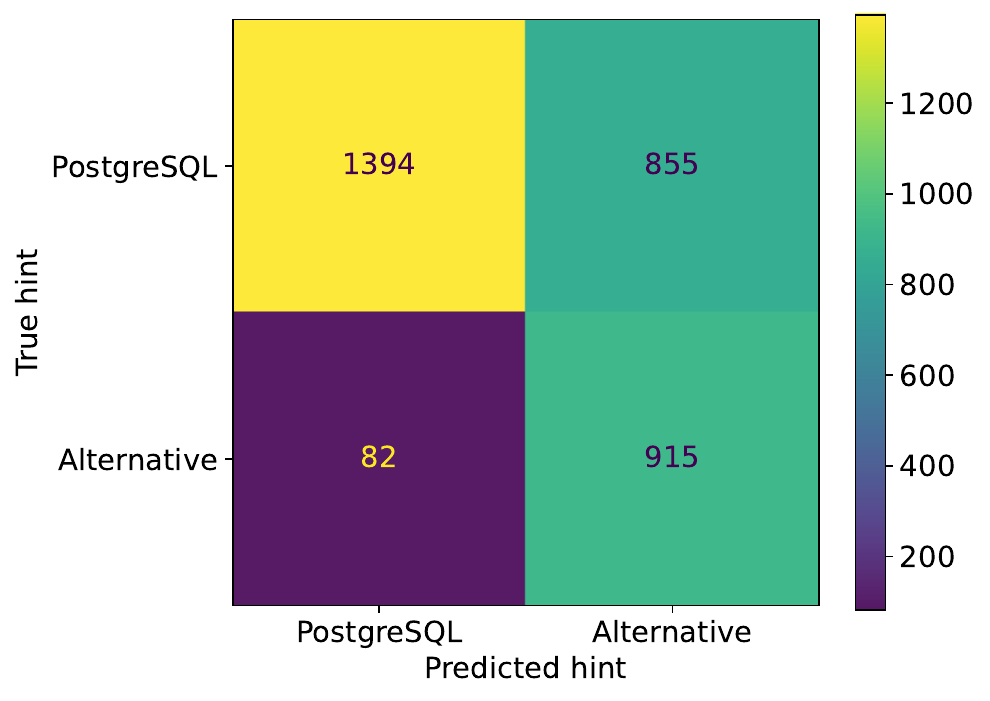}  

    \caption{Aggregate confusion matrix of \textsc{LLMSteer} across cross-validation testing workloads. }
    \label{fig:confusion}
\end{wrapfigure}

Total and P90 latency are generally accepted measures of performance within the database community, and are tied directly to the performance and optimization outcomes of the system, so we place a larger emphasis on the relative reduction in these metrics. However, we also assess the performance of \textsc{LLMSteer} on traditional classification metrics. The system achieves a mean recall of $0.9177$ ($\sigma=0.0202$), precision of $0.5174$ ($\sigma=0.0183$), and AUROC of $0.8515$ ($\sigma=0.0143$). The mean accuracy of \textsc{LLMSteer} is $0.7113$ ($\sigma=0.0205$), surpassing the performance of the naive classifier (i.e., the classifier that always selects the prior) marginally, which would have an accuracy of $0.6929$. Figure~\ref{fig:confusion} shows the aggregate confusion matrix of \textsc{LLMSteer}; true-positives occur where the true hint and predicted hint are both the alternative, while true-negatives occur where the true hint and predicted hint are both PostgreSQL. The system prioritizes minimizing false-negatives at the expense of a higher false-positive rate, resulting in improved performance due to the asymmetric cost associated with steering the QO incorrectly.

\begin{figure}[htbp] 
    \centering
        \begin{subfigure}[b]{0.32\linewidth} 
            \centering
            \includegraphics[width=\linewidth]{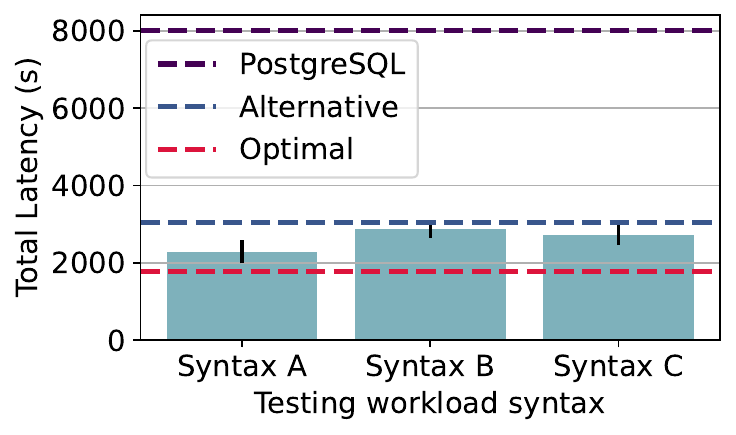}
            
            \caption{Trained on syntax A}
            \label{fig:syntaxA}
        \end{subfigure}
        \hspace{1pt}   % Space between the figures
        \begin{subfigure}[b]{0.32\linewidth}
            \centering
            \includegraphics[width=\linewidth]{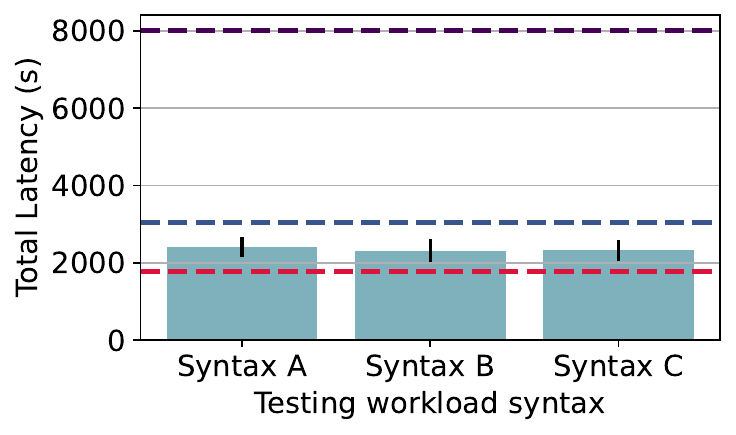}
            
            \caption{Trained on syntax B}
            \label{fig:syntaxB}
        \end{subfigure}
        \hspace{1pt}   % Space between the figures
        \begin{subfigure}[b]{0.32\linewidth}
            \centering
            \includegraphics[width=\linewidth]{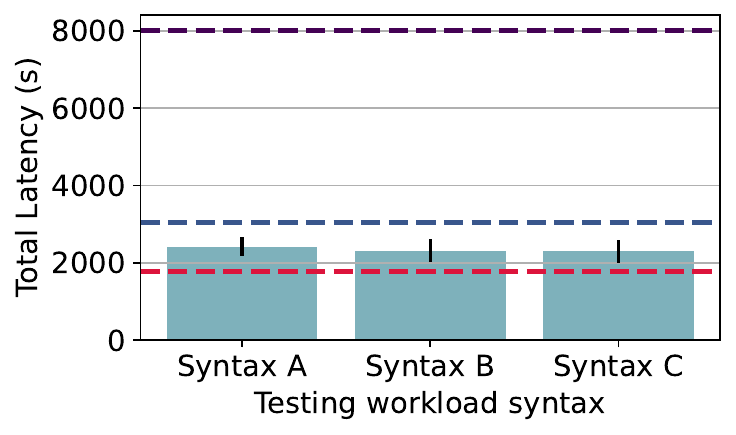}
            
            \caption{Trained on syntax C}
            \label{fig:syntaxC}
        \end{subfigure}
        
    \caption{Mean total latency of \textsc{LLMSteer} trained on augmented syntaxes across 10-fold cross-validation testing workloads. Syntax A represents original queries, Syntax B represents formatted queries with spaced indention, Syntax C represents formatted queries with tabbed indentation.}
    \label{fig:robustness}
\end{figure}

\paragraph{Experiment 2: Robustness to Syntactic Changes} SQL queries in original training and testing workloads are structured as single-line declarative statements. In practice, database users will rarely structure queries like this, as it impedes the ability to create complex queries and debug SQL statements. There are many ways to alter a query without changing its semantic meaning (Listings~\ref{lst:sql1} \& ~\ref{lst:sql2}), and LLMs are likely to produce different embeddings for queries based on their syntax. To assess robustness to such syntactic changes, we modified each query in various ways. We refer to "Syntax A" as the original phrasing of each query, and introduce "Syntax B" and "Syntax C," which use newline characters at the end of keyword blocks (i.e., \texttt{SELECT}, \texttt{FROM}, \texttt{WHERE}) and  use either spaces or tabs respectively for indentation. Figure~\ref{fig:robustness} shows that \textsc{LLMSteer} exhibits robustness to at least these classes of syntax changes. Notably, when \textsc{LLMSteer} trained on original queries (Syntax A), it was still effective on workloads with Syntax B and C; despite a 28\% increase in total latency when tested on syntax B and 27\% when tested on Syntax C, this still represented a reduction of 64\% relative to PostgreSQL. \textsc{LLMSteer} performed best when tested on a workload with the same syntax as it was trained on, but when trained on queries with Syntax B and C in particular, we observed minimal decrease in performance regardless of the syntax used in the testing workload.

\paragraph{Challenge: Scaling to more hints} Unfortunately, our simplified approach did not scale. Considering PostgreSQL's 48 hint sets, there are too few queries associated with each class, making it challenging for a classifier to learn the complex relationship between queries and hints. The distribution of queries across the collection of hints is also skewed, with the most frequently optimal hint set occurring 525x more often than the least frequently optimal. Despite this, even in the absence of a more complex strategy, the ability to steer the optimizer between just two alternatives leads to significantly improved performance.

\section{Conclusion and Future Work}\label{sec:conclusion}

In this extended abstract, we present \textsc{LLMSteer}, demonstrating its usage in effectively steering query optimizers. Benchmarked against PostgreSQL's default query optimizer, results from initial experimentation show that \textsc{LLMSteer} is capable of reducing total and tail latency by 72\% on average. We were surprised to discover that \textsc{LLMSteer} worked, since established wisdom of the database community indicates that the system should \textit{not} have been successful. With this, we have far more questions than answers.

\subsection{Reasons for Optimism}
We believe that \textsc{LLMSteer} opens up a number of future research directions.

\paragraph{The right embedding model?} The quality of embeddings are often highly dependent on the downstream task and the LLM used. At the time of this work, OpenAI's \texttt{text-embedding-3-large} did not rank within the top 30 models on the overall massive text embedding benchmark (MTEB)~(\cite{muennighoff2022mteb}). There may be models which can create richer representations of SQL queries, containing additional semantic information that may be helpful in steering optimizers. It is unclear whether open source embedding models with fewer parameters can be just as effective as their larger counterparts, and quantization techniques also present a promising alternative to using models with strictly fewer parameters. The effects of quantization on embedding quality are yet to be explored in the context of query optimization, but methods like \cite{frantar2023gptqaccurateposttrainingquantization}'s GPTQ, \cite{xiao2023smoothquant}'s SmoothQuant, and \cite{lin2024awq}'s AWQ may be essential to preserving performance gains in production systems. 

\paragraph{Fine-tune an LLM?} Given these preliminary results, it seems plausible that attaching a classification head to a language model and allowing the model to use its own activations in the selection of hints could be effective. More broadly, we question whether LLMs can be fine-tuned to perform the task of steering query optimizers? That is, can we teach an LLM to select the optimal hint given a query in a few-shot setting, or by fine-tuning an LLM on SQL directly, and would this prove to be more effective than \textsc{LLMSteer}? Further, what other features can we provide to an LLM to enhance performance on this task --- for example, could an LLM outperform existing methods if surfaced with a text representation of query plans?

\noindent\makebox[\textwidth][c]{
\begin{minipage}[t]{0.47\linewidth}
\begin{lstlisting}[language=sql, frame=single, numbers=none, caption={Example SQL query performing joins using the INNER JOIN keyword.}, label={lst:sql1}, xleftmargin=1pt, xrightmargin=1pt]
SELECT t.title AS movie, cn.country_code AS country
FROM title AS t INNER JOIN movie_companies AS mc ON t.id = mc.movie_id AND company_name AS cn INNER JOIN mc ON cn.id = mc.company_id
WHERE t.production_year > 2005;
\end{lstlisting}
    \end{minipage}
    \hspace{10pt}
    \begin{minipage}[t]{0.47\linewidth}
\begin{lstlisting}[language=sql, frame=single, numbers=none, caption={Example SQL query performing joins in the WHERE clause.}, label={lst:sql2}, xleftmargin=1pt, xrightmargin=1pt]
SELECT t.title AS movie, cn.country_code AS country
FROM company_name AS cn,
    movie_companies AS mc,
    title AS t
WHERE t.production_year > 2005
    AND t.id = mc.movie_id
    AND cn.id = mc.company_id;

    
\end{lstlisting}
    \end{minipage}
    \label{lst:sql}
} 

\subsection{Proceeding with Caution}

While we are cautiously optimistic about using LLMs for query optimization, we plan to investigate several additional possible explanations for \textsc{LLMSteer}'s surprising behavior.

\paragraph{Syntax} Additional investigation into how syntax impacts performance is necessary. For example, assessing the effects of comma-first notation, inclusion of comments, formatting of keywords and identifiers (e.g., lowercase, uppercase, title case), and any combination of these modifications on the ability for models to learn and generalize is of practical importance. Obviously, being robust to simple semantic-preserving reformulations of queries is critical for any real-world deployment.

\paragraph{Is the benchmark in the LLM's training set?} There are still multiple dimensions on which \textsc{LLMSteer} must be evaluated, providing sufficient cause for caution. It is unclear if the LLM has been exposed to the query benchmarks used in this work; to establish that \textsc{LLMSteer} has the ability to generalize, stronger evidence is needed to determine whether the LLM has trained on, and overfit to, these datasets. As a result, we question how \textsc{LLMSteer} might perform on novel SQL queries that are significantly different from existing datasets? Current large language models are ``internet scale'', making evaluation increasingly difficult --- the creation of new query benchmarks is nontrivial, and although beneficial to the database community, once a new benchmark is released, the next generation of LLMs may be trained on the data, confounding the results of future studies. This cycle presents a unique challenge with no clear solution. However, ablation studies and perturbation analysis may yield compelling results, providing key evidence that further validates the performance and generalizability of the system.

\paragraph{Taking a hint (or three)} Developing an understanding of any limitations related to scaling the system beyond two hints will be helpful in extending the utility of this system. In future work, performance gains must also be compared to a variety of query optimizers and current SOTA methods such as those using reinforcement learning. Latency of query processing is not captured in this work --- evaluating overhead and the overall latency impact in the critical path for queries is necessary to justify the use of embeddings.

We plan to investigate the above questions and more. We are actively seeking collaborators for this work, so please reach out if you are interested!

%----------------------------------------------------------

\clearpage
\bibliography{neurips,cites-long}

\begin{thebibliography}{17}
\providecommand{\natexlab}[1]{#1}
\providecommand{\url}[1]{\texttt{#1}}
\expandafter\ifx\csname urlstyle\endcsname\relax
  \providecommand{\doi}[1]{doi: #1}\else
  \providecommand{\doi}{doi: \begingroup \urlstyle{rm}\Url}\fi

\bibitem[Anneser et~al.(2023)Anneser, Tatbul, Cohen, Xu, Pandian, Laptev, and Marcus]{anneser2023autosteer}
C.~Anneser, N.~Tatbul, D.~Cohen, Z.~Xu, P.~Pandian, N.~Laptev, and R.~Marcus.
\newblock Autosteer: Learned query optimization for any sql database.
\newblock \emph{Proceedings of the VLDB Endowment}, 16\penalty0 (12):\penalty0 3515--3527, 2023.

\bibitem[Frantar et~al.(2023)Frantar, Ashkboos, Hoefler, and Alistarh]{frantar2023gptqaccurateposttrainingquantization}
E.~Frantar, S.~Ashkboos, T.~Hoefler, and D.~Alistarh.
\newblock Gptq: Accurate post-training quantization for generative pre-trained transformers, 2023.
\newblock URL \url{https://arxiv.org/abs/2210.17323}.

\bibitem[Giakoumakis and Galindo-Legaria(2008)]{testing_mssql}
L.~Giakoumakis and C.~A. Galindo-Legaria.
\newblock Testing {SQL} {Server}'s {Query} {Optimizer}: {Challenges}, {Techniques} and {Experiences}.
\newblock \emph{IEEE Data Eng. Bull.}, 31:\penalty0 36--43, 2008.

\bibitem[Graefe and McKenna(1993)]{volcano}
G.~Graefe and W.~J. McKenna.
\newblock The {Volcano} {Optimizer} {Generator}: {Extensibility} and {Efficient} {Search}.
\newblock In \emph{Proceedings of the {Ninth} {International} {Conference} on {Data} {Engineering}}, {ICDE} '93, pages 209--218, Washington, DC, USA, 1993. IEEE Computer Society.
\newblock ISBN 978-0-8186-3570-0.
\newblock URL \url{http://dl.acm.org/citation.cfm?id=645478.757691}.

\bibitem[Heinrich et~al.(2022)Heinrich, Luthra, Kornmayer, and Binnig]{Heinrich_2022}
R.~Heinrich, M.~Luthra, H.~Kornmayer, and C.~Binnig.
\newblock Zero-shot cost models for distributed stream processing.
\newblock In \emph{Proceedings of the 16th ACM International Conference on Distributed and Event-Based Systems}, DEBS ’22. ACM, June 2022.
\newblock \doi{10.1145/3524860.3539639}.
\newblock URL \url{http://dx.doi.org/10.1145/3524860.3539639}.

\bibitem[Kipf et~al.(2018)Kipf, Kipf, Radke, Leis, Boncz, and Kemper]{kipf2018learned}
A.~Kipf, T.~Kipf, B.~Radke, V.~Leis, P.~Boncz, and A.~Kemper.
\newblock Learned cardinalities: Estimating correlated joins with deep learning.
\newblock \emph{arXiv preprint arXiv:1809.00677}, 2018.

\bibitem[Leis et~al.(2015)Leis, Gubichev, Mirchev, Boncz, Kemper, and Neumann]{leis2015good}
V.~Leis, A.~Gubichev, A.~Mirchev, P.~Boncz, A.~Kemper, and T.~Neumann.
\newblock How good are query optimizers, really?
\newblock \emph{Proceedings of the VLDB Endowment}, 9\penalty0 (3):\penalty0 204--215, 2015.

\bibitem[Lin et~al.(2024)Lin, Tang, Tang, Yang, Chen, Wang, Xiao, Dang, Gan, and Han]{lin2024awq}
J.~Lin, J.~Tang, H.~Tang, S.~Yang, W.-M. Chen, W.-C. Wang, G.~Xiao, X.~Dang, C.~Gan, and S.~Han.
\newblock Awq: Activation-aware weight quantization for on-device llm compression and acceleration.
\newblock \emph{Proceedings of Machine Learning and Systems}, 6:\penalty0 87--100, 2024.

\bibitem[Marcus et~al.(2021)Marcus, Negi, Mao, Tatbul, Alizadeh, and Kraska]{marcusbao}
R.~Marcus, P.~Negi, H.~Mao, N.~Tatbul, M.~Alizadeh, and T.~Kraska.
\newblock Bao: Making learned query optimization practical.
\newblock In \emph{Proceedings of the 2021 International Conference on Management of Data}, SIGMOD '21, page 1275–1288, New York, NY, USA, 2021. Association for Computing Machinery.
\newblock ISBN 9781450383431.
\newblock \doi{10.1145/3448016.3452838}.
\newblock URL \url{https://doi.org/10.1145/3448016.3452838}.

\bibitem[Muennighoff et~al.(2022)Muennighoff, Tazi, Magne, and Reimers]{muennighoff2022mteb}
N.~Muennighoff, N.~Tazi, L.~Magne, and N.~Reimers.
\newblock Mteb: Massive text embedding benchmark.
\newblock \emph{arXiv preprint arXiv:2210.07316}, 2022.
\newblock \doi{10.48550/ARXIV.2210.07316}.
\newblock URL \url{https://arxiv.org/abs/2210.07316}.

\bibitem[Negi et~al.(2021)Negi, Marcus, Kipf, Mao, Tatbul, Kraska, and Alizadeh]{negi2021flow}
P.~Negi, R.~Marcus, A.~Kipf, H.~Mao, N.~Tatbul, T.~Kraska, and M.~Alizadeh.
\newblock Flow-loss: Learning cardinality estimates that matter.
\newblock \emph{arXiv preprint arXiv:2101.04964}, 2021.

\bibitem[Selinger et~al.(1979)Selinger, Astrahan, Chamberlin, Lorie, and Price]{systemr}
P.~G. Selinger, M.~M. Astrahan, D.~D. Chamberlin, R.~A. Lorie, and T.~G. Price.
\newblock Access {Path} {Selection} in a {Relational} {Database} {Management} {System}.
\newblock In J.~Mylopolous and M.~Brodie, editors, \emph{{SIGMOD} '79}, {SIGMOD} '79, pages 511--522, San Francisco (CA), 1979. Morgan Kaufmann.
\newblock ISBN 978-0-934613-53-8.
\newblock \doi{10.1016/B978-0-934613-53-8.50038-8}.
\newblock URL \url{https://www.sciencedirect.com/science/article/pii/B9780934613538500388}.

\bibitem[van Renen et~al.(2024)van Renen, Horn, Pfeil, Vaidya, Dong, Narayanaswamy, Liu, Saxena, Kipf, and Kraska]{redshift_workload}
A.~van Renen, D.~Horn, P.~Pfeil, K.~E. Vaidya, W.~Dong, M.~Narayanaswamy, Z.~Liu, G.~Saxena, A.~Kipf, and T.~Kraska.
\newblock Why {TPC} is not enough: {An} analysis of the {Amazon} {Redshift} fleet.
\newblock \emph{Proceedings of the VLDB Endowment}, 2024.
\newblock URL \url{https://www.amazon.science/publications/why-tpc-is-not-enough-an-analysis-of-the-amazon-redshift-fleet}.

\bibitem[Woltmann et~al.(2023)Woltmann, Thiessat, Hartmann, Habich, and Lehner]{fastgres}
L.~Woltmann, J.~Thiessat, C.~Hartmann, D.~Habich, and W.~Lehner.
\newblock {FASTgres}: {Making} {Learned} {Query} {Optimizer} {Hinting} {Effective}.
\newblock \emph{Proceedings of the VLDB Endowment}, 16\penalty0 (11):\penalty0 3310--3322, Aug. 2023.
\newblock ISSN 2150-8097.
\newblock \doi{10.14778/3611479.3611528}.
\newblock URL \url{https://dl.acm.org/doi/10.14778/3611479.3611528}.

\bibitem[Xiao et~al.(2023)Xiao, Lin, Seznec, Wu, Demouth, and Han]{xiao2023smoothquant}
G.~Xiao, J.~Lin, M.~Seznec, H.~Wu, J.~Demouth, and S.~Han.
\newblock Smoothquant: Accurate and efficient post-training quantization for large language models.
\newblock In \emph{International Conference on Machine Learning}, pages 38087--38099. PMLR, 2023.

\bibitem[Yi et~al.(2024)Yi, Tian, Ives, and Marcus]{10.1145/3663742.3663974}
Z.~Yi, Y.~Tian, Z.~G. Ives, and R.~Marcus.
\newblock Low rank approximation for learned query optimization.
\newblock In \emph{Proceedings of the Seventh International Workshop on Exploiting Artificial Intelligence Techniques for Data Management}, aiDM '24, New York, NY, USA, 2024. Association for Computing Machinery.
\newblock ISBN 9798400706806.
\newblock \doi{10.1145/3663742.3663974}.
\newblock URL \url{https://doi.org/10.1145/3663742.3663974}.

\bibitem[Zhu et~al.(2024)Zhu, Weng, Wei, Wu, Peng, Wang, Ding, Lian, Zheng, and Zhou]{pilotscope}
R.~Zhu, L.~Weng, W.~Wei, D.~Wu, J.~Peng, Y.~Wang, B.~Ding, D.~Lian, B.~Zheng, and J.~Zhou.
\newblock {PilotScope}: {Steering} {Databases} with {Machine} {Learning} {Drivers}.
\newblock \emph{PVLDB}, 17\penalty0 (5):\penalty0 980--993, 2024.
\newblock \doi{10.14778/3641204.3641209}.

\end{thebibliography}

%----------------------------------------------------------

\end{document}